\documentstyle[12pt]{article}

\newcommand{\beqn}{\begin{equation}}
\newcommand{\eeqn}{\end{equation}}

\begin{document}

\begin{titlepage}

\today          \hfill
\begin{center}
\hfill    LBL-40890	 \\
\hfill    UCB-PTH-97/50 \\

\setcounter{footnote}{1}
\vskip .25in
{\large \bf Path Integral Quantization of the Symplectic Leaves of 
the $SU(2)^{*}$\/ Poisson-Lie Group}\footnote{This work was supported in
part by the Director, Office of Energy Research, Office of High Energy and 
Nuclear Physics, Division of High Energy Physics of the U.S. Department of 
Energy under Contract DE-AC03-76SF00098 and in part by the National Science
Foundation under grant PHY-95-14797}
\vskip .25in

Bogdan Morariu
\footnote{email address: bogdan@physics.berkeley.edu}
\vskip .25in

%\author{Bogdan Morariu \thanks{email address: 
% bogdan@physics.berkeley.edu}}

{\em 	Department of Physics			\\
	University of California				\\
				and									\\
	Theoretical Physics Group			\\
	Lawrence Berkeley Laboratory	\\
	University of California				\\
	Berkeley, California 94720}
\end{center}
\vskip .25in

\begin{abstract}
The Feynman path integral is used to quantize the symplectic leaves of the 
Poisson-Lie group $SU(2)^{*}$\/. In this way we obtain the unitary 
representations of ${\cal U}_q(su(2))$. This is achieved by finding 
explicit Darboux coordinates and then using a phase space path integral. 
I discuss the $*$-structure of $SU(2)^{*}$\/ and give a detailed 
description of its leaves using various parametrizations. I also compare 
the results  with the path integral quantization of spin.
\end{abstract}
\end{titlepage}
%THIS PAGE (PAGE ii) CONTAINS THE LBL DISCLAIMER
%TEXT SHOULD BEGIN ON NEXT PAGE (PAGE 1)
\renewcommand{\thepage}{\roman{page}}
\setcounter{page}{2}
\mbox{ }

\vskip 1in

\begin{center}
{\bf Disclaimer}
\end{center}

\vskip .2in

\begin{scriptsize}
\begin{quotation}
This document was prepared as an account of work sponsored by the United
States Government. While this document is believed to contain correct
 information, neither the United States Government nor any agency
thereof, nor The Regents of the University of California, nor any of their
employees, makes any warranty, express or implied, or assumes any legal
liability or responsibility for the accuracy, completeness, or usefulness
of any information, apparatus, product, or process disclosed, or represents
that its use would not infringe privately owned rights.  Reference herein
to any specific commercial products process, or service by its trade name,
trademark, manufacturer, or otherwise, does not necessarily constitute or
imply its endorsement, recommendation, or favoring by the United States
Government or any agency thereof, or The Regents of the University of
California.  The views and opinions of authors expressed herein do not
necessarily state or reflect those of the United States Government or any
agency thereof, or The Regents of the University of California.
\end{quotation}
\end{scriptsize}

\vskip 2in

\begin{center}
\begin{small}
{\it Lawrence Berkeley Laboratory is an equal opportunity employer.}
\end{small}
\end{center}

\newpage
\renewcommand{\thepage}{\arabic{page}}

\setcounter{page}{1}

% main text is here

\section{Introduction}

The Feynman path integral reveals in a geometric intuitive way the relation
between classical and quantum dynamics. However there are few examples of
path integral quantizations on compact phase spaces. These are interesting 
because they have finite dimensional Hilbert spaces. The simplest example 
is a phase space with the topology of a torus. A more interesting case is 
obtained by considering a phase space with the topology of the sphere 
$S_2$\/. Quantization of this gives the spin. A path integral quantization 
is described in~\cite{NR,AFS}. Here I will present a generalization of 
this result, the case of the deformed spin.

Let $G$\/ be a Lie group. On the vector space $g^*$\/ dual 
to the Lie algebra $g$\/ of $G$\/ there is a natural Poisson structure.
In terms of linear coordinates $e_i$\/ and $f_{ij}^k$\/ the structure 
constants of the group it has the form
\[
\{ e_i,e_j \} = f_{ij}^k \, e_{k}
\]
and it is known as the Lie-Kirillov-Kostant Poisson bracket. Its symplectic 
leaves are the orbits of the coadjoint action~\cite{kirillov}. 
The quantization of this 
bracket is the universal enveloping algebra ${\cal U}(g)$\/ which is the 
associative algebra with generators $e_i$\/ and relations
\[
[e_i,e_j]=i\,\hbar \, f_{ij}^k \, e_k.
\]
Quantization of the coadjoint orbits of a Lie group $G$ gives its 
unitary representations~\cite{kirillov}. Various methods were used to
quantize these symplectic leaves  including geometric quantization and 
the Feynman path integral~\cite{NR,AFS}. Note that the vector space $g^*$\/ 
can be thought of as an abelian group. The above picture can be 
generalized to include Poisson brackets on non-abelian groups $G^*$\/
usually called the dual Poisson-Lie groups. This will be extensively 
discussed in Section~\ref{sec-dual}. Quantization of their symplectic leaves
gives the unitary representations of the quantum group ${\cal U}_q(g)$\/. 
This can be summarized in the picture below.
\[
\begin{array}{ccc}
	Fun(G^*)			&	\rightarrow	&	Fun_q(G^*) \cong {\cal U}_q(g)		\\
	\uparrow			& 					&	\uparrow					\\
	Fun(g^{*})		& \rightarrow	&	{\cal U}(g)	
\end{array}
\]
The  quantization axis is horizontal, with classical Poisson-Lie groups
on the left and their quantizations on the right. The vertical axis 
corresponds to deformation of the abelian case to the non-abelian case. 
Note that the abelian case can be obtained from the non-abelian case by 
looking at an infinitesimal neighborhood of the unit of the group, and 
rescaling coordinates appropriately. Throughout this paper I will refer to 
the lower part of the picture already discussed in~\cite{NR,AFS} as the 
trivial case\footnote{The Poisson bracket on $G$\/
discussed in Section~\ref{sec-dual} is trivial in this case}, and to 
the upper part  as the Poisson case.
 
I will use the Feynman path integral to quantize the symplectic
leaves of $SU(2)^*$\/. In doing this I will follow closely the method
used in~\cite{NR}. In fact, a strong parallel exists both at the classical
and the quantum levels. Classically, the leaves coincide in the trivial and 
Poisson cases once expressed in terms of Darboux coordinates. Consequently, 
at the quantum level we have the same Hilbert space and the two quantum
algebras are isomorphic. The path integral has the same form in the
trivial and Poisson cases, but one has to insert different functions
to obtain $su(2)$\/ or ${\cal U}_q(su(2))$\/ generators.

In Section~\ref{sec-dual}, I review some general Poisson-Lie theory mainly
to fix the notation and to list some results used later in the paper. The
results in this section are given using complex coordinates. In 
Section~\ref{sec-leaves}, I describe the reality structures of $SU(2)$\/,
its dual and its double. I also give a detailed description of the 
symplectic leaves of $SU(2)^*$\/. 

In Section~\ref{sec-piq}, I describe
Darboux coordinates, formulate the path integral and find the radius 
quantization condition using a quantization condition similar to~\cite{NR}. 
I also define the Hilbert space and obtain the matrix elements
of diagonal operators. In Section~\ref{sec-matrix}, I study  general
matrix elements and show that they are representations of the quantum 
group algebra. In the last section I draw some conclusions and
suggest how this work might be generalized. Finally, the appendix reviews
the isomorphism of $Fun_q(SU(2)^*)$\/ and ${\cal U}_q(su(2))$\/ and the 
derivation  of the Poisson bracket on $SU(2)^*$\/ from
$Fun_q(SU(2)^{*})$\/.

\section{Dual Pairs of Poisson-Lie Groups}

\label{sec-dual}

A {\em Poisson-Lie Group} (PLG) is a pair $(G,\{,\})$\/ where $G$\/ is a 
Lie group and $\{,\}$\/ is a Poisson bracket on $G$\/ which is compatible 
with the group operations of multiplication and inversion~\cite{Drinfeld}. 
The compatibility determines the Poisson structure at an arbitrary 
point from its values in the vicinity of the group unit. A PLG can be 
equivalently described as a {\em Poisson Hopf algebra}\/ $Fun(G)$\/ which
is a commutative Hopf algebra with a compatible Poisson algebra. In what 
follows I will freely exchange these two dual descriptions.

The Poisson bracket on the group determines a Lie algebra structure on the 
cotangent space $g^{*}$ of the Lie group. Let $h_{1}$ and $h_{2}$ be two 
functions on the group $G$. Then:
\[	 [dh_{1},dh_{2}]^{*} \equiv d\{h_{1},h_{2}\}	\]
defines a Lie algebra $(g^{*},[,]^{*})$\/. One can check that this 
definition is independent of the choice of functions used to represent 
cotangent vectors. Let $\{e_{i}\}$\/ be a basis of $g$\/, $\{e^{i}\}$\/ 
its dual basis in $g^{*}$\/,  and $f_{ij}^{k}$ and $\tilde{f}^{ab}_{c}$ 
the corresponding structure constants. The compatibility of the Poisson 
and group structures imposes restrictions on the two Lie algebras. In 
terms of the structure constants, they read
\beqn	f_{ij}^{s} \tilde{f}_{s}^{ab} - f_{is}^{a} \tilde{f}_{j}^{sb} +
	f_{is}^{b} \tilde{f}_{j}^{sa} - f_{js}^{b} \tilde{f}_{i}^{sa} +
	f_{js}^{a} \tilde{f}_{i}^{sb} = 0. \label{cocycle}
\eeqn
In fact, similarly to a Lie group being determined up to some global 
features by its Lie algebra, a PLG is in one to one correspondence with a 
{\em Lie bialgebra}\/ (LBA).  This is a pair $(g,g^{*})$\/ of Lie algebras 
dual as vector spaces whose structure constants satisfy (\ref{cocycle}). 
Note that the LBA structure is symmetric between $g$ and $g^*$\/, so to 
each LBA we can associate a pair of PLGs $G$ and $G^*$.

An equivalent definition of a LBA is given in terms of the cocommutator 
$\delta$ the dual of the $[,]^{*}$ commutator 
\[ \delta : g \rightarrow \wedge^{2} g, \, \langle \delta (x), \xi \wedge 
\eta \rangle =\langle x ,[\xi,\eta]^{*}\rangle,x\in g,\xi,\eta \in g^*.\]
Jacobi for $[,]^{*}$ implies co-Jacobi $(\delta \otimes id) \circ\delta=0$.
The compatibility condition~(\ref{cocycle}) translates into the cocycle 
condition
\[ \delta ([x,y])=[\Delta(x),\delta(y)]+[\delta(x),\Delta(y)]	\]
where $\Delta(x)=x\otimes 1 + 1\otimes x$\/ and similarly for $y$\/.

 A {\em quasi-triangular Lie bialgebra} is a LBA such that there exists a
 $r \in g \otimes g$\/ which, for all $x \in g$\/ satisfies:

\begin{enumerate}
	\item \( \delta(x)=[r, \Delta(x)]; \)
	\item \(I=r+\sigma(r) \) is adjoint invariant $[I,\Delta(x)]=0$\/. 
				Here $\sigma$ is the permutation operator;
	\item	\( (\delta \otimes id) r = [r_{13},r_{23}] \),\/ 
			\( (id \otimes \delta) r = [r_{13},r_{12}].	\)
\end{enumerate}

A {\em factorizable Lie bialgebra} is a quasi-triangular LBA such that $I$ 
is non-degenerate. One can use $I$ to identify $g$ and $g^{*}$. The 
factorization refers to the fact that any \( x \in g \) can be decomposed 
as \( x=x_{+} - x_{-} \)\/. Here 
\[	x_{+}=\langle r, \xi \otimes id\rangle,~
	x_{-}=-\langle r, id \otimes \xi \rangle \]
for some $\xi \in g^{*}$ satisfying \(x=\langle I,\xi\otimes id \rangle \)\/. 
Such a $\xi$\/ always exists since $I$ is non-degenerate.

A PLG $G$ is quasi-triangular if its tangent LBA $g$ is 
quasi-triangular. Similarly a PLG is factorizable if its tangent LBA is 
factorizable.

One can define two important Poisson brackets \( \{,\}_{\pm}\) on 
a quasi-triangular LBA.
\beqn
	\{f,h\}_{\pm} = \langle r,\nabla f \otimes \nabla h \rangle \pm
						\langle r,\nabla' f \otimes \nabla' h \rangle	
\eeqn
where
\[	\langle \nabla f(x),\xi\rangle \equiv \frac{d}{dt} f(e^{t \xi} x),~
\langle\nabla' f(x),\xi\rangle \equiv \frac{d}{dt} f(x e^{t \xi} )	\]
are the left and right gradients respectively.
The \( \{,\}_{-}\) Poisson bracket makes $G$ into a PLG. I will denote it
simply by $\{,\}$\/. The other bracket
\( \{,\}_{+}\) is also very important since it is non-degenerate almost 
everywhere and makes $G$ into a symplectic manifold.

For every representation $\rho$ one can explicitly write the Poisson 
relations 
for the matrix elements of $T(x)=\rho(x)$ which are coordinates on the 
group as
\beqn
	\{T_1,T_2 \} = [ r_{+} , T_1 T_2 ] \label{rTT} 
\eeqn
where $r_{+} = (\rho \otimes \rho) r$\/ and the 
subscript specifies the position in the tensor product. It is also useful
to define $r_{-} = - (\rho \otimes \rho) \sigma (r)$.\/ 

The standard example of a factorizable PLG is $SL(N,C)$\/. In this case
\[
r=\frac{1}{2} \sum_{i,j=1}^{N-1} (A^{-1})_{ij} \, H_i \otimes H_j +
\sum_{i < j} E_{ij} \otimes E_{ji}
\]
where $A$\/ is the Cartan matrix, $H_i$\/ are Cartan generators and 
$E_{ij}$\/ are generators which in the fundamental representation are
represented by matrices with only one non-vanishing entry equal to 
one in the $ij$\/ position. In this case we can give an explicit 
description of the dual group $SL(N,C)^{*}$ and its Poisson structure 
despite the fact that it is not quasi-triangular.
Let \(SL(N,C)^{*} \)\/ be the group of pairs of upper and lower triangular 
matrices 
\( \left\{ \left( L^{+},L^{-} \right) \right\} \)\/ where

\begin{equation}	L^{+} =\left( \begin{array}{ccc}
						a_1		&				&	 \ast		\\
									& \ddots		&				\\		
						0			&				&   a_n
						\end{array} \right),~~
	L^{-} =\left( \begin{array}{ccc}
						a_1^{-1}		&				&	 0			\\
										& \ddots		&				\\		
							\ast		&				&   a_n^{-1}
						\end{array} \right)	,~~
\prod_{i=1}^{N} a_{i} = 1.				\label{SLNC}		
\end{equation}
The group multiplication is given by multiplying corresponding matrices 
within each pair. Using the same notation for matrix group elements and 
functions on the group, the Poisson brackets are:
\begin{eqnarray}
\{L^{+}_{1},L^{+}_{2} \}=[r_{\pm},L^{+}_{1} L^{+}_{2}] , \nonumber \\
\{L^{-}_{1},L^{-}_{2} \}=[r_{\pm},L^{-}_{1} L^{-}_{2}],\label{PoissonLL}\\
\{L^{+}_{1},L^{-}_{2} \}=[r_{+},L^{+}_{1} L^{-}_{2} ]. \nonumber
\end{eqnarray}
One can also define 
\[
L= (L^{-})^{-1} L^{+} 
\]
and the Poisson brackets above become
\beqn \{ L_{1} , L_{2} \}  =  L_{1} r_{+} L_{2} + L_{2} r_{-} L_{1} -
										r_{+}L_{1} L_{2} -  L_{1} L_{2} r_{-} 
								. \label{LL}
\eeqn
The derivation of this bracket from the quantum commutation relations is
discussed in the appendix.
The map from $(L^+,L^-)$\/ to $L$\/ is not one to one. It is a
\( 2^{N-1} \) cover.  Later we will define reality structures on this 
Poisson algebras.

Now I will give a more detailed description of the $SL(2,C)$\/ and
 $SL(2,C)^{*}$\/ groups. Let
\[
T =  \left( \begin{array}{cc}  a & b \\ c & d 
				\end{array}	\right),~~
 L = \left( \begin{array}{cc} \alpha & \beta \\ \gamma & \delta 
				\end{array}	\right).	
\]
The classical r-matrices can be written as $4 \times 4$ matrices
\[ r_{+} = \left( \begin{array}{cccc}
		1/4 & 0 & 0 & 0 \\
		0 & -1/4 & 1 & 0 \\
		0 & 0 & -1/4 & 0 \\
		0 & 0 & 0 & 1/4
	\end{array}	\right) ,~~
    r_{-} = \left( \begin{array}{cccc}
		-1/4 & 0 & 0 & 0 \\
		0 & 1/4 & 0 & 0 \\
		0 &-1 & 1/4 & 0 \\
		0 & 0 & 0 & -1/4
	\end{array}	\right).
\]
Using (\ref{rTT}) after some algebra one obtains
\begin{eqnarray}	
	\{a,b\} & = & ab / 2 ,	\nonumber	 	\\
	\{a,c\} & = &   ac / 2	 , 	\label{TTPoisson} \\
	\{a,d\} & = &	cd	,	\nonumber	 \\
	\{b,c\} & = &	0	,	\nonumber	\\
	\{b,d\} & = &	bd / 2	,	\nonumber	 \\
	\{c,d\} & = &	cd / 2 	.	\nonumber
\end{eqnarray}
Similarly using (\ref{LL}) one obtains
\begin{eqnarray}	
	\{\alpha,\beta \} & = &	\alpha \beta 	,				\nonumber	 	\\
	\{\alpha,\gamma\} & = & - \alpha \gamma , 			\label{Poisson}\\
	\{\alpha,\delta\} & = &	 	0				,				\nonumber	 	\\
	\{\beta,	\gamma\} & = &	\alpha (\alpha-\delta) 	,\nonumber	 	\\
	\{\beta,	\delta\} & = &	\alpha	\beta 			,\nonumber	 	\\
	\{\gamma,\delta\} & = &	- \alpha	\gamma  			.	\nonumber
\end{eqnarray}

A further decomposition of $L^+$\/ as a diagonal matrix
and an upper diagonal matrix with unit entries on the diagonal, and 
of $L^-$\/ as a diagonal matrix and a lower diagonal matrix with unit 
entries on the diagonal, is possible. For the $SL(2,C)^{*}$\/ case, we have
\[ L^+ =\left( \begin{array}{cc} a & 0 \\ 0 & a^{-1}\end{array}\right)
		\left( \begin{array}{cc} 1 & \chi_{+} \\ 0 & 1 \end{array}\right)
   , ~~
 L^- =\left( \begin{array}{cc} a^{-1} & 0 \\ 0 & a\end{array}\right)
		\left( \begin{array}{cc} 1 &  0\\ -\chi_{-} & 1 \end{array}\right).
\]
It corresponds to Gauss's decomposition of $L$\/
\[		
L= \left( \begin{array}{cc} 1 &  0\\ \chi_{-} & 1 \end{array}\right)
	\left( \begin{array}{cc} a^{2} & 0 \\ 0 & a^{-2}\end{array}\right)
	\left( \begin{array}{cc} 1 & \chi_{+} \\ 0 & 1 \end{array}\right).
\]

To every LBA $(g,g^*)$\/ we can associate a factorizable LBA called the 
{\em double Lie bialgebra} $(d,d^{*})$.
First we define $d=g+g^{*}$\/,  i.e. the direct sum of vector spaces. 
It has a natural bilinear form $\langle,\rangle_d$\/ defined in terms of 
the dual pairing $\langle,\rangle$ of $g$ and $g^{*}$
\[
	\langle(x,\xi),(y,\eta)\rangle_{d} \equiv 
\langle x,\eta\rangle+\langle y,\xi\rangle, x,y \in g,\,\xi,\eta\in g^{*}.
\]
We define on $d$ the unique Lie algebra such that:
\begin{enumerate}
	\item $g$ and $g^{*}$ are subalgebras;
	\item	the bilinear form $\langle,\rangle_{d}$\/ determined by the dual 
			pairing is adjoint invariant.
\end{enumerate}
On the basis of $d$ given by $\{e_{i},e^{i} \} \),  the 
commutator \( [,]_d \)\/ has the form
\begin{quote}
	\(  [ e_i ,e_j ]_d = f_{ij}^{k} ~ e_{k}, \)

	\(  [ e^i ,e^j ]_d = f^{ij}_{k} ~ e^{k}, \)

	\(  [ e^i ,e_j ]_d = f_{ik}^{i} ~ e^{k} - \tilde{f}_{ik}^{j} ~ e_{k}.\)
\end{quote}
Also \( d^{*} \equiv g^{*} \oplus g \)\/, i.e.  it is the direct sum of 
Lie algebras $[e^i,e_j]_{d^*}=0$\/.
The pair \( (d,d^{*} ) \) is a factorizable LBA with 
\( r_d \equiv e^i \otimes e_i \in d\otimes d\), thus it is a projector on 
the $g$\/ factor. Note that $sl(N,C)$\/ is almost the double of one of its 
Borel subalgebras\footnote{It is the double of a Borel subalgebra divided 
by the Cartan subalgebra.}. We can exponentiate $d$ to a Lie group $D$ and 
$\{,\}_{-}$ will make it into a PGL. 

The simplest example of the above structure is obtained if we start from 
the trivial LBA $(g,g^*)$\/, i.e. $g$\/ is a Lie algebra and $g^*$\/ its 
dual with the trivial commutator. $G$ is a Lie group with Lie algebra $g$ 
and  $G^*= g^*$\/ is an abelian group. $D$ is the cotangent bundle 
$T^* G= G\times g^*$\/.  The $\{,\}_{+}$\/ bracket is the canonical 
Poisson bracket on the cotangent bundle, and $\{,\}_{-}$ is the Lie bracket
on $g^*$ extended by left translations to the cotangent bundle.

The double $D$ of a factorizable PLG $G$ can be described in more detail. 
As a group it is isomorphic with $G\times G$\footnote{This is only true for
complex groups. If $G$\/ has a reality structure the double is obtain by 
imposing a reality structure on $G^c\times G^c$\/ where $G^c$\/ is the
complexification of $G$\/.}. The groups $G$ and $G^*$ are 
subgroups of $D$ and are embedded as follows
\[	G \subset G\times G, ~~ T \rightarrow (T,T),	\]
\[ G^* \subset G \times G, ~~ L \rightarrow (L^{+},L^{-}).	\]
Almost all elements $(x,y)$\/ of the double  can be written in factorized 
form
\begin{equation}
		(x,y) =  (T,T)^{-1} (L^+,L^-)	= 
(\tilde{L}^{+},\tilde{L}^{-})^{-1} (\tilde{T},\tilde{T}). \label{xTL}
\end{equation}

A pair of Poisson manifolds $(P,P')$\/ is called a 
{\em dual pair}\/~\cite{LuW,STS2} if there exists a symplectic manifold 
$S$\/ and two projections $\pi$\/ and $\pi'$\/

\[
\begin{array}{rcccl}
		&				&	S	&				&			\\
\pi	& \swarrow	&	 	& \searrow	&	\pi'	\\
P	&				&		&				& P'
\end{array}
\]
such that the sets of functions which are pullbacks of functions on $P$\/ 
and  $P'$\/ centralize each other
\[
\{\pi^*(f),\pi'^{*}(f') \}_S = 0,
\]
An important theorem~\cite{LuW,STS} 
states that each symplectic leaf of $P$\/ is obtained by projecting 
on $P$\/ the preimage of an element $a$\/ of $P'$\/
\[
\pi ( \pi'^{-1} (a)), ~~ a \in P' .
\]

The manifolds $D/G$\/ and $G\setminus D$\/ form a dual pair. The symplectic 
manifold is the double $D$\/ of $G$ with the $\{,\}_{+}$ bracket. 
The following projections
\[
\begin{array}{rcccl}
					&				&	D	&				&			\\
\pi				& \swarrow	&	 	& \searrow	&	\pi'	\\
G \setminus D	&				&		&				& D/G	
\end{array}
\]
can be used to induce  Poisson structures on $D/G$\/ and $G\setminus D$\/. 
Since $D$\/ is factorizable $G^* \cong G\setminus D$\/. Moreover the 
Poisson structure
induced on $G\setminus D$\/ from $D$\/ coincides with the original Poisson
structure on $G^*$\/. Then the above theorem gives the symplectic leaves 
of $G^*$\/. In particular if $G$\/ is 
factorizable, $\pi'(x,y) \equiv x y^{-1} = a$\/
and the preimage of $a$ has elements of the form $(ay,y)$\/. Then  
$\pi(x,y) = y^{-1} x = y^{-1} a y$\/, thus the symplectic leaves are 
given by the orbits of the coadjoint action of $G$ on $G\setminus D$\/. 
This action is also known as the dressing action~\cite{STS} 
\[
		G \times (G\setminus D) \rightarrow G\setminus D, ~ 
				(y,a) \rightarrow  y^{-1}ay.
\]

\section{Symplectic Leaves}
\label{sec-leaves}
In the first part of this section, I will discuss the $SL(N,C)$\/ case. So 
far, everything was complex. The simplest reality structure one can impose 
is to require everything to be real. We then obtain $SL(N,R)$\/, its
double, dual etc. However,  we want to obtain $SU(N)$\/. We start on the 
double with the reality structure 
\[			x^{\dag} = y^{-1}.	\]
Since $G$\/ and $G^*$\/ are subgroups, this induces the following reality 
structures
\beqn	T^{\dag} = T^{-1}, ~ (L^{+})^{\dag} = (L^{-})^{-1}.\label{reality}
\eeqn
Once we impose (\ref{reality}) the dual group is no longer simply connected,
since $a_i$\/ in (\ref{SLNC}) are real and non-zero. Define $SU(N)^{*}$ as 
the component 
connected to the unit element of the group.
\[	
SU(N)^{*} = 
\{ (L^+,L^-) \in SL^*(N,C) \mid (L^{+})^{\dag}=(L^{-})^{-1},~ a_i > 0 \}.
\]
We can also describe $SU(N)^*$\/ in terms of $L$ as the set of hermitian, 
positive definite matrices of determinant one. Then the map 
\(   (L^+,L^-) \rightarrow L = (L^-)^{-1}L^{+}  \)\/ is one to one and the 
factorization is unique.

 For $SU(2)^{*}$  the reality structure is $\bar{\alpha} = \alpha$\/, 
$\bar{\delta} = \delta$\/, 
$\bar{\beta} = \gamma$\/. 

To summarize, the double of $SU(N)$\/ is $SL(N,C)$\/, and the factorization
(\ref{xTL}) can be written \( x=T^{-1} L^{+}\)\/, that is to say, any matrix
of determinant one can be 
decomposed uniquely as the product of a special unitary matrix and  an 
upper triangular matrix with real positive diagonal entries\footnote{Note 
that $y$\/ is not independent $y=(x^{\dag})^{-1}$}. 

In particular the double of $SU(2)$\/ is the proper Lorentz group 
$SL(2,C)$\/. It is interesting  to note that the double of the 
trivial PLG $SU(2)$, i.e. its cotangent bundle, is the proper homogeneous 
Galilean group.

Using the two factorizations
\[	(x,y)=(T^{-1} L^+ ,T^{-1} L^-)=((\tilde{L}^{+})^{-1}\tilde{T},
(\tilde{L}^{-})^{-1}\tilde{T})	\]
and the projections
\(	\pi(x,y)= y^{-1}x\)\/, \( \pi'(x,y)= x y^{-1}	\) we obtain the 
following form for the symplectic leaves
\[ \pi( \pi'^{-1}((\tilde{L}^{+})^{-1} \tilde{L}^{-}))=
					\{ (L^{-})^{-1} L^{+}
	= (\tilde{T})^{-1} \tilde{L}^{-} (\tilde{L}^{+})^{-1} \tilde{T} |
	\tilde{T} \in SU(2) \}
\]
where \( (\tilde{L}^{+},\tilde{L}^{-}) \in SU(2)^{*}\)
is fixed, and $\tilde{T}$ parametrizes the leave. This is just the orbit 
of the right Poisson coadjoint action of $SU(2)$\/ on $SU(2)^{*}$\/
\[
L \rightarrow T^{-1} L T .
\]

It is convenient to use an exponential parametrization of 
$L=(L^{-})^{-1} L^{+}$
\[
L = \exp(x_i\sigma_i)  = 
		\cosh(r)+\sinh(r)
\left( \begin{array}{cc} n_3 & n_{-} \\ n_{+} &-n_3\end{array}\right)
\]
where $\sigma_i$\/'s are the Pauli matrices, $r^2=\sum_{i} x_i^2$\/ and 
$n_i=x_i/r$\/.
Since $tr(L) = 2 \cosh(r)$\/ is invariant under the coadjoint action we 
see that the simplectic leaves are spheres of radius $r$\/ except for the
$r=0$\/ leaf, which is zero dimensional.
In terms of the exponential parametrization, the Poisson 
algebra~(\ref{Poisson}) becomes

\[	\{x_{\pm}, x_{3}\}=\pm x_{\pm} (x_{3}+r\coth(r)),	\]
\[	\{ x_{-}, x_{+} \} = 2 x_{3} (x_{3}+r\coth(r)).	\]
Since $r$\/ is constant on symplectic leaves it must be central in the 
above Poisson algebra, which can be checked by direct computation.
These Poisson spheres and their quantization were first studied 
in~\cite{Podles}.
One can parametrize the radius $r$\/ sphere using stereographic 
projection coordinates $z,\bar{z}$
\[ z=\frac{x_{-}}{r-x_{3}}, ~
 \bar{z}=\frac{x_{+}}{r-x_{3}}.	\]
After some straightforward algebra we obtain
\[
\{ \bar{z}, z\}_{r} = \frac{1}{2} \left(1+ z \bar{z} \right)^{2}
\left(\frac{z \bar{z}-1}{z \bar{z}+1} + \coth(r) \right).
\]
The right action of $SU(2)$\/ on $z$\/ by fractional transformations 
\[
z'= \frac{\bar{a}z-b}{\bar{b}z+a}
\]
is a Poisson action i.e. $a,b,c,d$\/ have non-trivial bracket given by
 (\ref{TTPoisson}). Since our path integral is formulated in real time, 
we do a Wick rotation and obtain the Minkowski Poisson bracket
\beqn
\{ \bar{z}, z\}_{r} = \frac{i}{2}\left(1+ z\bar{z}\right)^{2}
\left(\frac{z \bar{z}-1}{z \bar{z}+1}+\coth(r) \right) \label{zzq}
\eeqn
differing from the original one by a phase factor. 

Using non-singular coordinates around the south pole $w=-1/z$\/ the Poisson
bracket becomes
\[
\{ \bar{w}, w\}_{r}= \frac{i}{2} \left(1+ w \bar{w}\right)^{2}
\left(-\frac{w \bar{w}-1}{w \bar{w}+1}+ \coth(r)\right)
\]
thus the Poisson structure is not north-south symmetric.
The infinite $r$\/ limit is singular at the south pole. This particular 
Poisson  structure and its quantization was studied in~\cite{BZ,BZ1}.

The small $r$\/ limit is dominated by the $\coth(r)$\/ term and
\beqn
\{ \bar{z}, z\}_{r} \approx \frac{i}{2} \coth(r)
\left(1+ z \bar{z} \right)^{2}.		\label{zzc}
\eeqn
This is the standard Poisson bracket on a sphere of radius 
$\coth^{1/2}(r)$\/. The right action by fractional transformations 
on (\ref{zzc}) leaves this Poisson bracket invariant. Thus the small radius 
symplectic leaves are almost rotationally invariant.

Next we obtain the symplectic form on the leaves. Let $f,h$\/ be functions 
on the leaf; each $f$\/ defines a vector field 
$v_f$ such that $v_f(h) =\{f,h\}$\/. Then the symplectic form is defined by
\[
\Omega (v_f,v_h) \equiv  \{h,f\}.
\]

In local coordinates, the Poisson bracket and the symplectic form have 
the form
\[		\{f,h\}= P^{ij} \, \partial_i f \,
 \partial_j h	,	~	
\Omega = \frac{1}{2} \Omega_{ij} dx^{i}\wedge dx^{j},	\]
and the two antisymmetric tensors satisfy
\[		P^{ij} \Omega_{jk}   = \delta_{k}^{i}	. \]
In complex coordinates, this is simply 
\(P^{\bar{z} z}\Omega_{z \bar{z}}=1\),
and gives
\[
	\Omega = -\frac{2}{i}
\frac{ \bar{dz} \wedge dz}{(1+z \bar{z})^2}
\left(\frac{z \bar{z}-1}{z \bar{z}+1}  + \coth(r)\right)^{-1}
 =-\frac{\Omega_{0}}{n_3 + \coth(r)  } ,
\]
where $\Omega_{0}$\/ is the standard area 2-form on the unit sphere.

\section{Path Integral Quantization}
\label{sec-piq}

The path integral quantization of the Poisson algebra on the leaves of 
$su(2)^*$\/ was discussed 
in~\cite{NR,AFS}. Quantization of these leaves gives the unitary 
representations of $SU(2)$\/. We will do the same for the 
symplectic leaves
above and obtain the unitary representations of 
\({\cal U}_{q} (su(2)) \)\/
algebra. This is in fact a Hopf algebra but we concentrate here on 
the algebra structure\footnote{The coproduct and antipode of the 
$L^{\pm}$\/
generators are the same as in the classical Poisson-Hopf algebra}.

Before starting the quantization we have to find canonical coordinates on 
the leaves. Note that 
\[ \Omega_0 = \sin \theta \, d\theta \wedge d\phi = 
 d (-\cos (\theta)) \wedge d\phi	\]
thus $(-\cos(\theta),\phi)$\/ are Darboux coordinates on the 
standard $S_2$\/.  Similarly
\[ \Omega = d [-\ln (n_3+\coth (r))] \wedge d\phi \] 
so we define
\[ J \equiv -\ln \left[
 \frac{n_3+\coth (r)}{(\coth^2 (r) - 1)^{1/2}} \right]
= - \ln \left[ \cosh(r) + \sinh(r) \, n_3 \right]   \]
where the denominator was fixed by the requirement that $J$\/ spans a 
symmetric interval $(-r,r)$. We have 
\(\Omega = dJ \wedge d\phi = d( J\,d\phi) \)\/ so we define the Poincare 
1-form $\Theta$\/
\[	\Theta = J\, d\phi + c \, d\phi	\]
where c is a constant to be fixed later. Thus the Poisson sphere of radius
$r$\/ is parametrized by $J$\/ and $\phi$\/ as
\[n_3 = \sinh^{-1}(r) (e^{-J} - \cosh(r)), ~
n_{\pm}=(1-n_3^2)^{1/2}~ e^{\mp i \phi}.
\]

The Poisson algebra on  any leaf can be quantized, but in general
these quantum algebras will not have unitary representations. Unitarity 
leads to a  quantization of the radius of the Poisson sphere. Before 
starting the Poisson case let us review two different quantization 
conditions used in~\cite{NR,AFS} for the trivial case.
In~\cite{AFS} a geometric quantization condition similar to that used for 
the Dirac monopole or the Wess-Zumino-Witten model was used to obtain the 
allowed values of the radius. 
The action must be continuous as the path crosses over the poles. 
Equivalently 
\beqn
e^{i/ \hbar \oint \Theta} = 1		\label{quantAFS}
\eeqn
where the integral is 
over an infinitesimal loop around the poles. However this condition was 
only used to determine the characters of the representations. Also note 
that, unlike the Dirac monopole  where the action 
is a configuration space action, 
both in the trivial and the Poisson case one has a phase space action.

However in~\cite{NR} it was shown that in order to obtain the matrix 
elements of 
$su(2)$\/ a non-trivial phase has to exist as the path crosses the poles.
Requiring the correct matrix elements one obtains the quantization 
condition
\beqn
	e^{i/\hbar \oint \Theta} = -1	\label{quantization}
\eeqn
This gives the same result as (\ref{quantAFS}) for the Cartan generator 
and thus for the 
characters. Here I will use~(\ref{quantization}) and show that 
we obtain the standard matrix elements of the quantum qroup generators.
 
Imposing~(\ref{quantization}) at the north and south poles we obtain the 
quantization $r = N\hbar /2$\/ where $N$\/ is a positive integer. For 
$N$\/ odd one can set $c=0$\/ but a non-zero $c$ is 
required for even $N$\/.
The simplest choice is $c= \hbar /2$\/. We can write 
the two cases together as
\[ \Theta = (J + M \hbar/2)\, d \phi, ~ M=0,1.	\]

Next I list some of the functions on the Poisson sphere that I will 
quantize, expressed in terms of Darboux variables $J,\phi$ 
\begin{eqnarray}	
\alpha 	&=& e^{-J} \nonumber	 													\\
\beta &=&(-1+2\cosh(r)e^{-J}-e^{-2J})^{1/2}e^{i\phi}\label{s2functions} \\
\gamma 	&=&(-1+2\cosh(r)e^{-J}-e^{-2J})^{1/2}e^{-i\phi}  \nonumber	\\
\delta 	&=&2\cosh(r) - e^{-J}	 \nonumber									\\
	a 		&=& e^{-J/2}						\nonumber							\\
\chi_{\pm} &=&(-1+2\cosh(r)e^{J}-e^{2J})^{1/2}e^{\pm i\phi} \nonumber 	
\end{eqnarray}
The general structure of this functions is
\[ {\cal O} (J,\phi) = {\cal F}(J) e^{i p \phi}, ~ p=0,\pm 1.	\]
Note also that
\[	tr(L) = 2 \cosh(r) = 2 \cosh \left( N \hbar/2 \right) =
	q^N + q^{-N},	\]
where we introduced $q \equiv e^{ \hbar /2}$\/. Since $tr(L)$\/ only 
depends on $r$\/, it is central in the Poisson algebra and will be 
central in the quantum algebra. In fact $tr(L)$\/ is the Casimir of 
\({\cal U}_{q} (su(2)) \)\/.

Next we discuss the Feynman path integral. Consider first for 
simplicity a Hamiltonian $H(J)$\/, i.e. a function of $J$\/ and not 
of $\phi$\/.
Wave functions are  functions on $S_1$\/  (or periodic functions 
of $\phi$\/) and let $\mid \phi \rangle $ be a $\phi$\/ eigenvector.
The propagator on $S_1$\/ can be expressed in terms of the propagator on 
the covering space of $S_1$,  which is the real line by
\beqn
 \langle \phi' \mid e^{-\frac{i}{\hbar}H T} \mid \phi \rangle =
\sum_{n \in Z} 
\langle \phi' + 2\pi n \mid e^{-\frac{i}{\hbar} H T}\mid \phi\rangle_{0}
\eeqn
where formally
\beqn
\langle \phi'\mid e^{-\frac{i}{\hbar}H T}\mid \phi\rangle_{0} =
\int\!\!\int	\frac{{\cal D} J \, {\cal D} \phi}{2 \pi \hbar}  ~
e^{\frac{i}{\hbar} \int_{0}^{T} [\Theta - H(J) \, dt]}
\eeqn
where $\phi$\/ is integrated over the whole real line and $J$\/ over the
$(-r,r)$\/ interval. To make sense of the formal expression we
divide $T$\/ into $P$\/ intervals and let $\phi_0 = \phi, \phi_P=\phi'$\/.
Then
\begin{equation}
\langle \phi'\mid e^{-\frac{i}{\hbar}H T}\mid \phi\rangle_{0} =
\int \frac{\prod_{i} dJ_i}{2 \pi \hbar}
\int  \prod_{i} d\phi_i ~
e^{ i/ \hbar \sum_i [(J_i+c)(\phi_i-\phi_{i-1}) - H(J_i) T/P]} \label{PI}
\end{equation}
The $\phi$\/ integration can be performed leading to delta functions which
allow us to do all but one of the $J$\/ integrals. Then the propagator on 
$S_1$\/ takes the form
\[
\langle \phi'\mid e^{-\frac{i}{\hbar}H T}\mid \phi\rangle =
\sum_{n\in Z}\int_{- N \hbar/2}^{N \hbar/2}\frac{dJ}{2\pi\hbar} ~
e^{-i/\hbar H(J)T}
e^{ i/\hbar(J+c)(\phi'+2\pi n) }
e^{ -i/\hbar (J+c)\phi }
\]
Using the Poisson resummation formula
\[
\sum_{n\in Z} e^{2\pi i n  \alpha} = \sum_{k\in Z} \delta(\alpha - k)
\]
we perform the last integral and obtain
\[
\langle \phi'\mid e^{-\frac{i}{\hbar}H T}\mid \phi\rangle =
\sum_{k \atop{ | J_k | \leq N\hbar/2}} 
% \sum_{k} 
\frac{e^{i k \phi'}}{\sqrt{2\pi}} ~
e^{-i/\hbar H(J_k)T}~
\frac{e^{- i k \phi}}{\sqrt{2\pi}}
\]
where $J_k = \hbar(k - M/2)$\/. The sum is over all integers 
$k$ such that $(-N+M)/2 \leq k \leq (N+M)/2$\/.
We see that not all states propagate. We can make 
the path integral unitary
by projecting out the states that do not propagate. 
Define the Hilbert space as the vector space spanned by the vectors
\[	
\mid m \rangle = \int\frac{d\phi}{\sqrt{2\pi}} ~ e^{i(m+M/2)\phi} 
\mid \phi \rangle , ~~  m=-j, \ldots,j
\]
where, according to angular momentum conventions, $j$\/ is a half integer 
such that $N=2j+1$ \/. 
Note that the exponent is always an integer and $N$\/ is the 
total number of
states. The maximum value $J=\pm N \hbar/2$\/ is not reached quantum 
mechanically. It differs from the results in~\cite{AFS} but agrees 
with~\cite{NR} as previously mentioned. It was 
pointed out in~\cite{NR} that
this is similar to the non-zero ground state energy of the harmonic 
oscillator. 

\section{Matrix Elements and the Quantum Algebra}
\label{sec-matrix}

Since this is a phase space path integral some care must be taken when 
quantizing functions which depend on canonically conjugate variables.
The standard mid-point prescription for a function of the form
${\cal J}(J){\Phi}(\phi)$\/ is to write it as 
${\cal J}(J_i){\Phi}[(\phi_i+\phi_{i-1})/2]$\/ in the path integral. Thus 
for functions of the form \( {\cal O} (J,\phi)={\cal F}(J) e^{ i p \phi} \) 
I will use ${\cal F}(J_i) e^{i p (\phi_i+\phi_{i-1})/2}$\/. To calculate the
matrix elements of such an operator we insert it in the path integral 
(\ref{PI}) with $H=0$\/ and take $T$\/ infinitesimal. For the prescription 
above it is sufficient to consider only one time interval. The matrix 
elements are
\[
\langle \phi' \mid {\cal O} \mid \phi \rangle =
\sum_{n\in Z} \int \frac{dJ}{2\pi\hbar} ~ 
e^{i/\hbar (J+c)(\phi'+2\pi n -\phi)}
{\cal F}(J) 
e^{i p (\phi'+2\pi n + \phi)/2} =
\]
\[
\sum_k \frac{e^{i k \phi'}}{\sqrt{2\pi}}~
				{\cal F}(J_k)~
		\frac{e^{-i (k-p) \phi}}{\sqrt{2\pi}}
\]
where \(J_k = \hbar(k-M/2-p/2) \)\/, and I used Poisson resummation 
before performing the $J$\/ integral. Then the matrix elements in 
the \( \{ \mid m \rangle \} \)\/ basis are given by
\beqn
({\cal O})_{m'm} = 
\langle m' \mid {\cal O} \mid m \rangle =
{\cal F}[(m'-p/2)\hbar] \, \delta_{m'-p-m,0}, ~ m=-j,\ldots,j.
\label{matrix}
\eeqn

Using the opposite mid-point prescription  
\( {\cal F} [(J_i+J_{i-1})/2] e^{ip\phi_i} \) gives the 
same matrix elements.
However in this case one has to consider at least two time intervals if 
working in the $\phi$\/ representation. This prescription is 
more convenient
when working in the $J$\/ representation.

We can use~(\ref{matrix}) to calculate matrix elements of 
any function on $SU(2)^*$\/. Mid-point prescription in the path integral 
results in a special ordering of the quantum operators, when expressed 
in terms of $J$\/ and $\phi$\/, called Weyl ordering. If one starts from 
the Gauss's decomposition, uses path integral to 
obtain the matrix elements 
of  $a$\/ and $\chi_{\pm}$\/ and then uses them to express $L^{\pm}$\/ as 
products of quantum matrices, we obtain the quantum commutation 
relations~\cite{FRT}.
Using~(\ref{matrix}) we obtain 
\begin{eqnarray}	
(a)_{m'm} & = & e^{- \hbar m'/2}\, \delta_{m'-m,0}  \label{Lmatrix}	\\
(\chi_{\pm})_{m'm} &=& 
(-1+2\cosh(\hbar(j+1/2))e^{\hbar(m' \mp 1/2)}-e^{2\hbar(m' \mp 1/2)})^{1/2}
						\, \delta_{m'-m \mp 1,0} 		 \nonumber
\end{eqnarray}
One can check by direct calculation that relations~(\ref{Lmatrix}) are 
representations of the algebra generated by $a,\chi_{\pm}$\/ with 
relations
\begin{eqnarray}
	\chi_{+} a &=& q 			a \chi_{+}					\nonumber\\
	\chi_{-} a &=& q^{-1}  	a \chi_{-}					\label{algebra}\\
	q\chi_{+}\chi_{-} -q^{-1}\chi_{-}\chi_{+}&=
												&\lambda(a^{-4}-1)	\nonumber
\end{eqnarray}
where $\lambda \equiv q-q^{-1}$\/. Using this we define the 
quantum matrices $L^{\pm}$ as
\[
L^{+}	=
\left( \begin{array}{cc}	a	&	0	\\
											0	&	a^{-1}	\end{array}	\right)
\left( \begin{array}{cc} 1 & \chi_{+} \\
								 0 & 1 \end{array}	\right),
\]
\[
L^{-}	=
\left( \begin{array}{cc}	a^{-1}	&	0	\\
											0	&	a	\end{array}	\right)
\left( \begin{array}{cc} 1 & 0 \\	
								 -\chi_{-} & 1 \end{array}	\right).
\]
One can use~(\ref{algebra}) to check that $L^{\pm}$\/ satisfies the 
quantum  group commutations relations~\cite{Drinfeld,FRT,Zumino} 
\begin{eqnarray}	
R_{\pm} L_{1}^{\pm} L_{2}^{\pm}=
L_{2}^{\pm} L_{1}^{\pm} R_{\pm} \nonumber	\\
R_{+} L_{1}^{+} L_{2}^{-} =
 L_{2}^{-} L_{1}^{+} R_{+} 		\label{RLL}	\\
R_{-} L_{1}^{-} L_{2}^{+} =
 L_{2}^{+} L_{1}^{-} R_{-} 		\nonumber
\end{eqnarray}
where the quantum matrices are given in the appendix.
Alternatively, using the representations 
\beqn		L^{+} = \left(	\begin{array}{cc}
								q^{-H/2} & q^{-1/2} \lambda X_{+}	\\
								0			& q^{H/2}
							\end{array}
						\right) ,~
		L^{-} = \left(	\begin{array}{cc}
								q^{H/2} 						& 	0	\\
								-q^{1/2} \lambda X_{-}	& q^{-H/2}
							\end{array}
						\right)	\label{Lplus}
\eeqn
of the quantum $L^{\pm}$\/ in terms of Jimbo-Drinfeld generators discussed 
in the appendix, the
relations~(\ref{algebra}) are equivalent to
\beqn
[H,X_{\pm} ]= \pm 2 X_{\pm}, ~ [X_+,X_-]=\frac{q^{H}-q^{-H}}{q-q^{-1}}
\eeqn

The Jimbo-Drinfeld generators of ${\cal U}_q (su(2))$\/ can be 
obtained in  the path integral by inserting
\begin{eqnarray}
H &=&  \hbar^{-1} \, 2J			\label{qgen}	\\
X_{\pm} &=& 
\lambda^{-1} [2(\cosh(r) -\cosh(J)]^{1/2} \, e^{\pm i\phi}.\nonumber 
\end{eqnarray}
Note that unlike $a$\/ and $\chi_{\pm}$\/ the insertions above are already
quantum. In addition while the functional dependence 
in terms of $J$\/ and $\phi$
can be easily obtained from (\ref{Lplus}) the overall normalization of
$X_{\pm}$\/ has been adjusted to give the standard result. 
The same kind of
normalization adjustments are necessary if one tries to insert the matrix
elements of $L^{\pm}$\/ directly into the path 
integral. This just reflects
ordering ambiguities of quantum operators. Alternatively one could get the
standard result without any adjustments of normalization 
by using a non-midpoint
prescription. For example  the off-diagonal element of $L^{+}$\/ equals
$a \chi_{+}$\/ with this specific ordering in the quantum case. Since the 
path integral gives time ordering we can obtain the 
desired quantum ordering
by using the following prescription
\[
e^{-J_{i}/2} 
(-1 + 2 cosh(r) e^{(J_i + J_{i-1})/2} - e^{J_i + 
J_{i-1}})^{1/2} e^{+ i\phi}
\]
Note that I only used a mid-point prescription for $\chi_{+}$\/
and not for $a$\/. The matrix elements obtained using~(\ref{matrix}) are
\begin{eqnarray}
(H)_{m'm} &=& 2 m \, \delta_{m'-m,0}  ,\nonumber  \\
(X_{\pm})_{m'm} &=& \{2\coth[\hbar(j + 1/2)]- 
 2\coth[\hbar(m \pm 1/2)]\}^{1/2}  \, \delta_{m'-m \mp 1,0} . \nonumber
\end{eqnarray}

The generators of $su(2)$\/ are obtained using
\begin{eqnarray}
\tilde{H} &=& 2 J, ~	\label{cgen}				 \\
\tilde{X}_{\pm} &=& (r^2 - J^2)^{1/2} \, e^{\pm i \phi}. \nonumber
\end{eqnarray}
In this case it is possible to 
write all generators without using $\hbar$\/ while in the deformed case a
different rescaling for each generator is required to eliminate $\hbar$\/.
The matrix elements obtained using~(\ref{matrix}) 
\begin{eqnarray}
(\tilde{H})_{m'm} &=& 2\hbar m \, \delta_{m'-m,0} , ~ \nonumber \\
(\tilde{X}_{\pm})_{m'm} &=& \hbar  \nonumber
 [(j+1/2)^{2}-(m \pm 1/2)^{2}]^{1/2}\, \delta_{m'-m \mp 1,0} 
\end{eqnarray}
are just the standard matrix elements of the $su(2)$\/ algebra
\[
	[\tilde{H},\tilde{X}_{\pm} ]= \pm 2\hbar \tilde{X}_{\pm}, ~ 
	[\tilde{X}_{+},\tilde{X}_{-}]= \hbar \tilde{H}.
\]
\section{Concluding Remarks}
In addition to trying to generalize the results in~\cite{NR,AFS} my goal
in this paper was to better understand the quantization (\ref{RLL}) of the 
Poisson bracket (\ref{PoissonLL}). Any $R_{\pm}$\/ satisfying
$R_{\pm} = 1 + \hbar r_{\pm} + {\cal O} ( \hbar^{2} ) $ used in (\ref{RLL})
would give the same Poisson bracket in the classical limit. The
${\cal O} ( \hbar^{2} ) $\/ and higher order terms are fixed by  
requireing that (\ref{RLL}) are commutation relations of a Hopf algebra 
deformation of the original Poisson-Hopf algebra. It is natural then to ask 
what is the relation of this quantization to the quantization known as Weyl 
quantization. Of course this 
question could be answered using algebraic methods without appealing to path 
integrals. At least for the case of $SU(2)$\/, I found that the functions 
$\chi_{\pm}$\/ and $a$\/ appearing in the Gauss's decomposition play a 
special role. Their quantization using Weyl ordering gives the same 
commutation relations as in the quantum group quantization. It would be 
interesting to investigate if this result still holds for an arbitrary 
$SU(N)$\/.

It should be possible to generalize the path integral formulated in 
this paper to arbitrary classical groups.
The similarity between the trivial and the Poisson cases for $SU(2)$\/
suggests that a starting point could be the path integral quantization of 
the coadjoint orbits of classical groups discussed  in~\cite{AFS}.  
 
The existence of a non-trivial phase as the path crosses the poles 
discussed in~\cite{NR} is present in the Poisson case too. A better 
understanding of the origin of this phase would be welcomed.

Let us now compare the trivial and Poisson cases. The symplectic
leaves in both cases are spheres parametrized by $(z,\bar{z})$\/ in 
stereographic 
projection. The group $SU(2)$\/ acts in the same way on the leaves 
in the two cases, i.e.  by standard rotations of the spheres, but in the
trivial case the bracket is invariant under the action, while in the
Poisson case the action is only a Poisson action. However, once the 
symplectic form is expressed in Darboux coordinates $(J,\phi)$\/ the
leaves appear to be identical. As a consequence the path integral
has the same form as in~\cite{NR,AFS}, but since the  
transformation to the 
Darboux variables is non-trivial in the Poisson case, $SU(2)$\/ acts
in a complicated way on the leaves, and functions on $SU(2)^*$\/ have a 
complicated dependence on $(J,\phi)$\/. Compare for example (\ref{qgen}) 
and (\ref{cgen}). Thus the same path integral generates
different matrix elements because we insert different functions in the
trivial and Poisson cases. This shows explicitly 
that on the same symplectic 
manifold  one can implement both a trivial and a Poisson symmetry. The
question  of which is the actual symmetry of the 
system is a dynamical one, 
and can only be answered after we know the Hamiltonian.  
Finally, I conjecture 
that as in the $SU(2)$\/ case, for an arbitrary classical group, the 
path integral has the same form in the trivial and Poisson cases.

\section*{Acknowledgements}

I would like to thank Professor Bruno Zumino for many useful discussions
and valuabe comments. I would also like to thank Paolo Aschieri and
Harold Steinacker for valuable input. This work was supported in part by 
the Director, Office of Energy Research, Office of High Energy and Nuclear
Physics, Division of High Energy Physics of the U.S. Department of Energy
under Contract DE-AC03-76SF00098 and in part by the National Science 
Foundation under grant PHY-90-14797.

\appendix

\section*{Appendix}

Here we list some relations defining the 
quantum group $Fun_q(SU(2)^{*})$\/ and discuss its relation to 
${\cal U}_q(su(2))$\/ \cite{Drinfeld,FRT,Zumino}.  We only discuss the 
algebra and ignore all other issues. The quantum 
qroup $Fun_q(SU(2)^{*})$\/ is a factorizable quasi-triangular Hopf algebra.
As an algebra it is generated by triangular matrices $L^{\pm}$\/ satisfying
quantum commutation relations
\begin{eqnarray}	
R_{\pm} L_{1}^{\pm} L_{2}^{\pm}=
L_{2}^{\pm} L_{1}^{\pm} R_{\pm} \nonumber	\\
R_{+} L_{1}^{+} L_{2}^{-} = L_{2}^{-} L_{1}^{+} R_{+} \label{RPLUSLL} \\
R_{-} L_{1}^{-} L_{2}^{+} = L_{2}^{+} L_{1}^{-} R_{-} \nonumber
\end{eqnarray}
where
\[ R_+ = q^{-1/2} \left( \begin{array}{cccc}
									q & 0 & 0 & 0 \\
									0 & 1 & \lambda & 0 \\
									0 & 0 & 1 & 0 \\
									0 & 0 & 0 & q
								\end{array}
							\right), ~
R_- = q^{1/2} \left( \begin{array}{cccc}
									q^{-1} & 0 & 0 & 0 \\
									0 & 1 & 0 & 0 \\
									0 & -\lambda & 1 & 0 \\
									0 & 0 & 0 & q^{-1}
								\end{array}
							\right) .
\]
The universal enveloping algebra ${\cal U}_q(su(2))$ is a quasi-triangular
Hopf algebra. It has generators
$H,X_{\pm}$\/ which satisfy the Jimbo-Drinfeld relations
\beqn
[H,X_{\pm} ]= \pm 2 X_{\pm}, ~ [X_+,X_-]=\frac{q^{H}-q^{-H}}{q-q^{-1}}.
\eeqn
In~\cite{FRT} it was shown that these two Hopf algebras are isomorphic. The
isomorphism is given by
\beqn		L^{+} = \left(	\begin{array}{cc}
								q^{-H/2} & q^{-1/2} \lambda X_{+}	\\
								0			& q^{H/2}
							\end{array}
						\right) ,~
		L^{-} = \left(	\begin{array}{cc}
								q^{H/2} 						& 	0	\\
								-q^{1/2} \lambda X_{-}	& q^{-H/2}
							\end{array}
						\right)	.
\eeqn
As in the classical case we can define the matrix $L = (L^{-})^{-1}L^{+} $. 
It satisfies the following equation:
\beqn
	R_{+}^{-1} L_{1} R_{+} L_{2} = L_{2} R_{-}^{-1} L_{1} R_{-}	\label{RLRL}
\eeqn
as can be checked using (\ref{RLL}).

In the classical limit we define  $r_{\pm}$ matrices 
by $R_{\pm} = 1 + \hbar r_{\pm} + {\cal O} ( \hbar^{2} ) $. Then
\[	(1 - \hbar r_{+}) L_{1} (1 + \hbar r_{+}) L_{2} =
	L_{2} (1 - \hbar r_{-}) L_{1} (1 + \hbar r_{-}) + {\cal O} (\hbar^{2}) 
\]
and we obtain the following Poisson structure
\[ \{ L_{1} , L_{2} \} \equiv \lim_{\hbar \rightarrow 0} 
\frac{ [ L_{1} , L_{2} ] }{ - \hbar} =
	+ L_{1} r_{+} L_{2} + L_{2} r_{-} L_{1}
	-	r_{+}L_{1} L_{2} - L_{1} L_{2} r_{-} .
\]
This is just the original Poisson bracket (\ref{LL}) which was the 
starting point for the path integral quantization.


\begin{thebibliography}{9}

\bibitem{NR} H. B. Nielsen, D. Rohrlich, {\em  A Path Integral to 
Quantize Spin}, Nucl. Phys. B299 (1988) 471-483

\bibitem{AFS} A, Alekseev, L. Faddeev, S. Shatashvili, {\em  Quantization 
of Symplectic Orbits of Compact Lie Groups by Means of the Functional 
Integral}, JGP. Vol. 5, nr.~3 (1989) 391-406

\bibitem{RS} N. Yu. Reshetikhin, M. A. Semenov-Tian-Shansky, {\em  Quantum 
R-matrices and Factorization Problems}, JGP. Vol. 5, nr.~4 (1988) 534-550

\bibitem{STS} M. A. Semenov-Tian-Shansky, {\em Dressing Transformations and 
Poisson Group Actions}, Publ. RIMS, Kyoto Univ.~21 (1985) 1237-1260

\bibitem{STS2} M. A. Semenov-Tian-Shansky, {\em Poisson-Lie Groups, Quantum 
Duality Principle and the Twisted Quantum Double}, Theor. Math. Phys. 93, 
nr.~2 (1992) 302-329

\bibitem{BZ} C. S. Chu, P. M. Ho, B. Zumino, {\em The Quantum 2-sphere as a
Complex Quantum Manifold}, Zf. Physik C70 (1996), 339; 
Preprint q-alg/9504003, April 1995

\bibitem{BZ1} C. S. Chu, P. M. Ho, B. Zumino, {\em The Braided Quantum 
2-Sphere}, Mod. Phys. Lett. A11 (1996) 307-316; 
Preprint q-alg/9507013, July 1995

\bibitem{Drinfeld} V. G. Drinfeld, {\em Quantum Groups}, ICM MSRI, Berkeley 
(1986) 798-820
 
\bibitem{FRT} L.D. Faddeev, N. Yu. Reshetikhin, L. A. Takhtajan 
{\em Quantization of Lie Groups and Lie Algebras}, Alg. i Anal. 1 (1989) 178

\bibitem{Zumino} B. Zumino, {\em Introduction to the Differential Geometry
of Quantum Groups}, K.Schm\"{u}dgen (Ed.), Math. Phys. X, Proc. X-th IAMP 
Conf. Leipzig (1990), Springer-Verlag (1990)

\bibitem{Podles} P. Podle\'{s}, {\em Quantum Spheres}, Lett. Math. Phys. 14 
(1987) 193

\bibitem{LuW} J. H. Lu, A. Weinstein, {\em Poisson-Lie Groups, Dressing 
transformations and Bruhat Decompositions}, J. Diff. Geom. 31 (1990) 510 

\bibitem{kirillov} A. A. Kirillov, {\em Elements of the Theory of 
Representation.}, Berlin, Heidleberg, New York: Springer 1976

\end{thebibliography}
\end{document}